\documentclass[]{spie}  

\usepackage{amsmath,amsfonts,amssymb}
\usepackage{graphicx}
\usepackage{wrapfig}
\usepackage{subcaption}
\usepackage[export]{adjustbox}
\usepackage[colorlinks=true, allcolors=blue]{hyperref}

\title{Optomechanical design of an ADC and image rotator for the HJST Coudé Spectrograph at McDonald Observatory}

\author[a]{Devika K Divakar}
\author[a]{Andrew Pinckney}
\author[a]{Joseph Strubhar}
\author[a]{Phillip MacQueen}
\affil[a]{McDonald Observatory, The University of Texas at Austin, 2515 Speedway Blvd, Texas 78712, USA}

\authorinfo{Further author information: \\a. Devika Divakar: devika.divakar@austin.utexas.edu\\  b. Phillip MacQueen: E-mail: pjm@astro.as.utexas.edu}

\begin{document} 
\maketitle

\begin{abstract}
High-resolution spectroscopic observations require stable slit illumination and mitigation of atmospheric refraction effects that degrade throughput and spectral fidelity. We present the optical and mechanical design of a Rotational Atmospheric Dispersion Corrector (RADC) for the coudé Tull Spectrograph on the Harlan J. Smith 2.7\,m Telescope. Atmospheric dispersion introduces wavelength-dependent image displacement that increases with zenith distance and reaches approximately 4 arcsec across the 350--1000 nm wavelength range at a zenith angle of 65$^\circ$, leading to wavelength-dependent slit losses and non-uniform slit illumination. The RADC employs a pair of counter-rotating Amici prisms to provide continuous correction over the operational zenith-angle range while maintaining image quality and compatibility with the existing coudé pre-slit optics. The optical and mechanical design of the deployable ADC subsystem, its integration within the limited relay envelope, and its expected impact on throughput and slit illumination stability are presented. In addition, we discuss a K-mirror image rotator to compensate for the field rotation at the coud\'e focus. This potential future upgrade will allow observations of extended objects and targets in confused fields by maintaining a fixed slit position angle on the sky regardless of the target's hour angle. Preliminary optical studies of the image-rotation concept and its integration considerations within the HJST coud\'e optical path are briefly summarized. Together, these developments support enhanced observing capabilities for high-resolution spectroscopy on the HJST.

\end{abstract}


\section{INTRODUCTION}
\label{sec:intro}  

High-resolution optical spectroscopy  plays a critical role in observational astrophysics, enabling detailed investigations of the composition, kinematics, and physical conditions of stars, planetary systems, and the interstellar medium. These applications require stable slit illumination, high throughput, and minimal wavelength-dependent systematics to preserve spectral fidelity over long integrations and through board spectral bandpasses. Atmospheric effects and field rotation can introduce variations at the spectrograph entrance slit that directly impact spectral stability and observing efficiency, particularly for broadband observations and extended targets. The 2.7 m Harlan J. Smith Telescope (HJST) at McDonald Observatory is a major facility for high-resolution spectroscopy and feeds the Tull Spectrograph (formerly 2dcoude), a cross-dispersed white-pupil echelle spectrograph designed for broad wavelength coverage and high spectral resolution \cite{Tull_1995}. The key telescope and site characteristics relevant to this work are summarized in Table~\ref{tab:hjst_specs}. The spectrograph operates through the coud\'e optical path of the telescope, where the incoming beam is redirected by a sequence of fold mirrors into a stationary spectrograph room located beneath the telescope floor \cite{TS2Guide}. This configuration provides a mechanically stable environment for precision spectroscopy while maintaining a fixed f/32.32 beam at the spectrograph entrance. The Tull Spectrograph has been extensively used for stellar abundance studies, radial velocity programs, and high-resolution spectroscopy over the visible wavelength range \cite{Endl_2012,Sneden_2009,Hackshaw_2025}. The key optical and operational characteristics of the Tull Spectrograph that drive the requirements of the current upgrade are summarized in Table~\ref{tab:ts_specs}. The existing optical train does not include atmospheric dispersion correction or field derotation capability. As telescope pointing moves away from the zenith, atmospheric refraction introduces wavelength-dependent image displacement at the slit plane. Across the operational wavelength range of 350--1000 nm, the uncorrected atmospheric dispersion can reach approximately 4 \,arcsec at a zenith angle of 65$^\circ$, significantly exceeding the typical slit widths used for high-resolution spectroscopy. This effect produces chromatic elongation of the stellar image, reducing throughput and introducing wavelength-dependent slit losses away from the guiding wavelength. In addition, image rotation during telescope tracking changes the orientation of the image at the slit plane during long integrations. 

\begin{table*}[ht]
\centering
\vspace{-5pt}
\caption{\footnotesize Key specifications of the Harlan J. Smith Telescope.}
\label{tab:hjst_specs}
\footnotesize
\begin{tabular}{|c|c|}
\hline
\textbf{Parameter} & \textbf{Specification} \\
\hline
Telescope & Harlan J. Smith Telescope (HJST) \\
Observatory & McDonald Observatory, University of Texas at Austin \\
Primary mirror diameter & 2.718 m  \\
Optical design & Ritchey - Chr\'etien reflector (at f/8.8)\\
Mount type & Equatorial mount  \\
Geographic latitude/longitude & 30.68$^\circ$ N/104.02$^\circ$ W \\
Observatory altitude & $\sim$ 2070 m  \\
Cassegrain focus characteristics & f/8.8 (straight and 3 bent cass ports) \\
Coud\'e focus characteristics & 5 mirror, f/32.32 \\

\hline
\end{tabular}
\end{table*}

\begin{figure}[ht]
\centering
\vspace{-15pt}
\includegraphics[width=\linewidth]{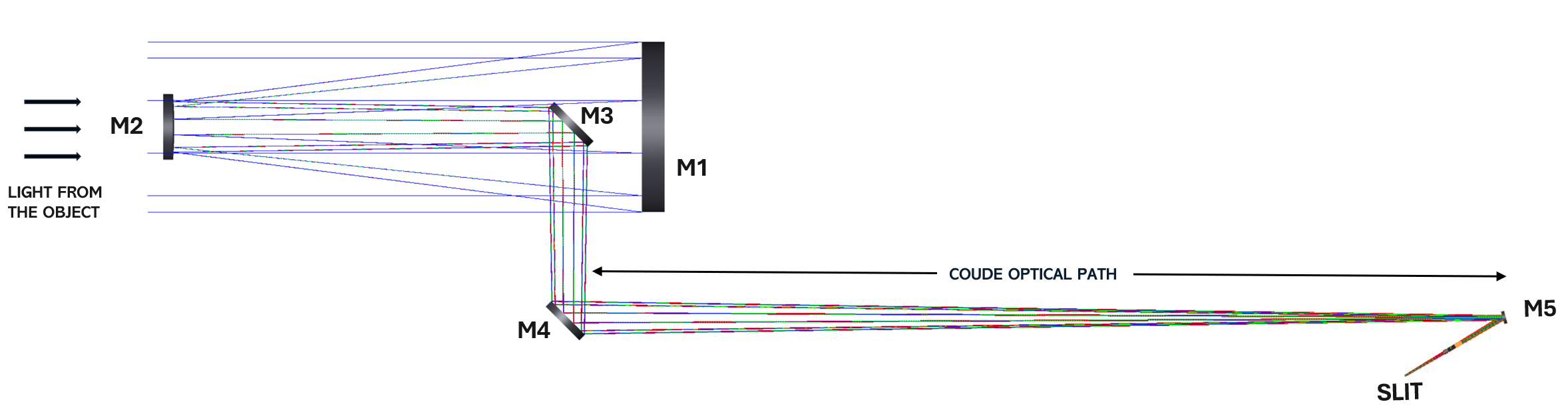}
\caption{\footnotesize
Optical layout of the upgraded HJST coud\'e optical path incorporating the RADC and K-mirror image rotator. The telescope beam is relayed through fold mirrors M1 - M5 while preserving the existing coud\'e beam geometry. The RADC provides broadband atmospheric dispersion correction using counter-rotating Amici prisms, while the reflective K-mirror assembly enables continuous image derotation and stabilization of the instrument position angle at the slit surface. The image rotator and ADC are between M5 and slit.
}
\label{fig:hjst}
\end{figure}
 \begin{table*}[ht]
\centering
\caption{\footnotesize Key specifications of Tull Spectrograph relevant to the upgraded optical system \cite{Tull_1995}.}
\label{tab:ts_specs}
\footnotesize
\begin{tabular}{|c|c|}
\hline
\textbf{Parameter} & \textbf{Specification} \\
\hline
Instrument & Tull Spectrograph\\
Image scale at the coud\'e focus &  2.348 arcsec mm$^{-1}$ or 426 \textmu m arcsec$^{-1} $ \\
Number of focii & Two (TSF1 and TSF3) \\
Maximum spectral resolving power & R $\sim$ 200,000 (TSF1) and R $\sim$ 60,000 (TSF3)\\
Echelle grating options & E1 (79.01 lines mm$^{-1}$,  $\theta_B$ = 63.43\textdegree)\,\\ &  E2 (52.67 lines mm$^{-1}$,  $\theta_B$ = 65.29\textdegree)\\
Slit dimensions for R $\sim$ 60,000 & $1.18'' \times 8.0''$ \\
Typical slit width range (TSF3) & 1.18$''$ - 2.36$''$ \\
Spectral coverage (TSF3-E2) & 350--1000 nm \\
Detector (Current) & Tektronix 2048 $\times$ 2048 CCD (TK3) \\
Detector (post 2026) & E2V 4096 $\times$ 4112, 15 um pixels \\
TK3 pixel size & 24 $\textmu $m  \\

\hline
\end{tabular}
\end{table*}
\noindent Figure~\ref{fig:hjst} illustrates the proposed upgraded HJST coud\'e  optical layout incorporating both the ADC and the image rotator. The incoming telescope beam is redirected through a sequence of fold mirrors (M3 - M5) while preserving the existing optical path and mechanical constraints of the spectrograph slit bench. In this paper, we present the design of an Atmospheric Dispersion Corrector (ADC) and an image rotator for the Tull Spectrograph. Together, these subsystems are intended to improve slit illumination stability, increase broadband throughput, maintain stable image orientation, and enhance the overall observing efficiency of the Tull Spectrograph for high-resolution spectroscopy applications.

\section{OPTICAL DESIGN}
\begin{figure}[ht]
\centering
\includegraphics[width=\linewidth]{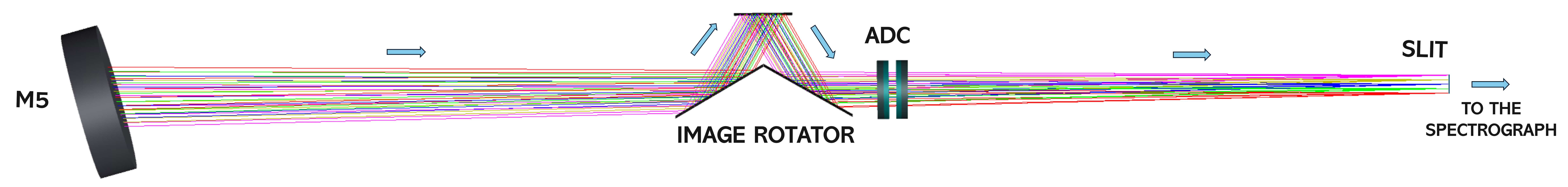}
\caption{\footnotesize
Optical layout of the upgraded HJST coud\'e optical path incorporating the K-mirror image rotator and RADC upstream of the Tull spectrograph slit. The incoming telescope beam from fold mirror M5 is directed through the image rotator assembly to compensate field rotation and maintain a fixed image orientation at the slit plane. The beam then passes through the counter-rotating Amici prism RADC, which provides broadband atmospheric dispersion correction over the 350--1000 nm wavelength range before entering the Tull Spectrograph slit.
}
\vspace{-10pt}
\label{fig:kmirror_radc_layout}
\end{figure}

The proposed upgrade places the image rotator and the ADC in the HJST coud\'e relay upstream of the Tull spectrograph slit. 
The upgraded system is designed for broadband operation over 350--1000 nm with an ADC field of view of approximately 2\,arcmin and an image rotator field of view of approximately 1\,arcmin. 
As shown in Figure~\ref{fig:kmirror_radc_layout}, the beam from fold mirror M5 first enters the image rotator assembly, which provides continuous image derotation while preserving broadband throughput. The beam then passes through the ADC, where wavelength-dependent atmospheric dispersion is corrected before the image is delivered to the spectrograph slit. This arrangement allows field rotation and atmospheric dispersion to be corrected before slit injection, improving slit illumination stability and reducing wavelength-dependent throughput losses. The two subsystems are functionally independent and can be operated separately. 

\subsection{IMAGE ROTATOR}
\begin{wrapfigure}{r}{0.35\textwidth}
\vspace{-10pt}
\centering
\includegraphics[width=0.9\linewidth]{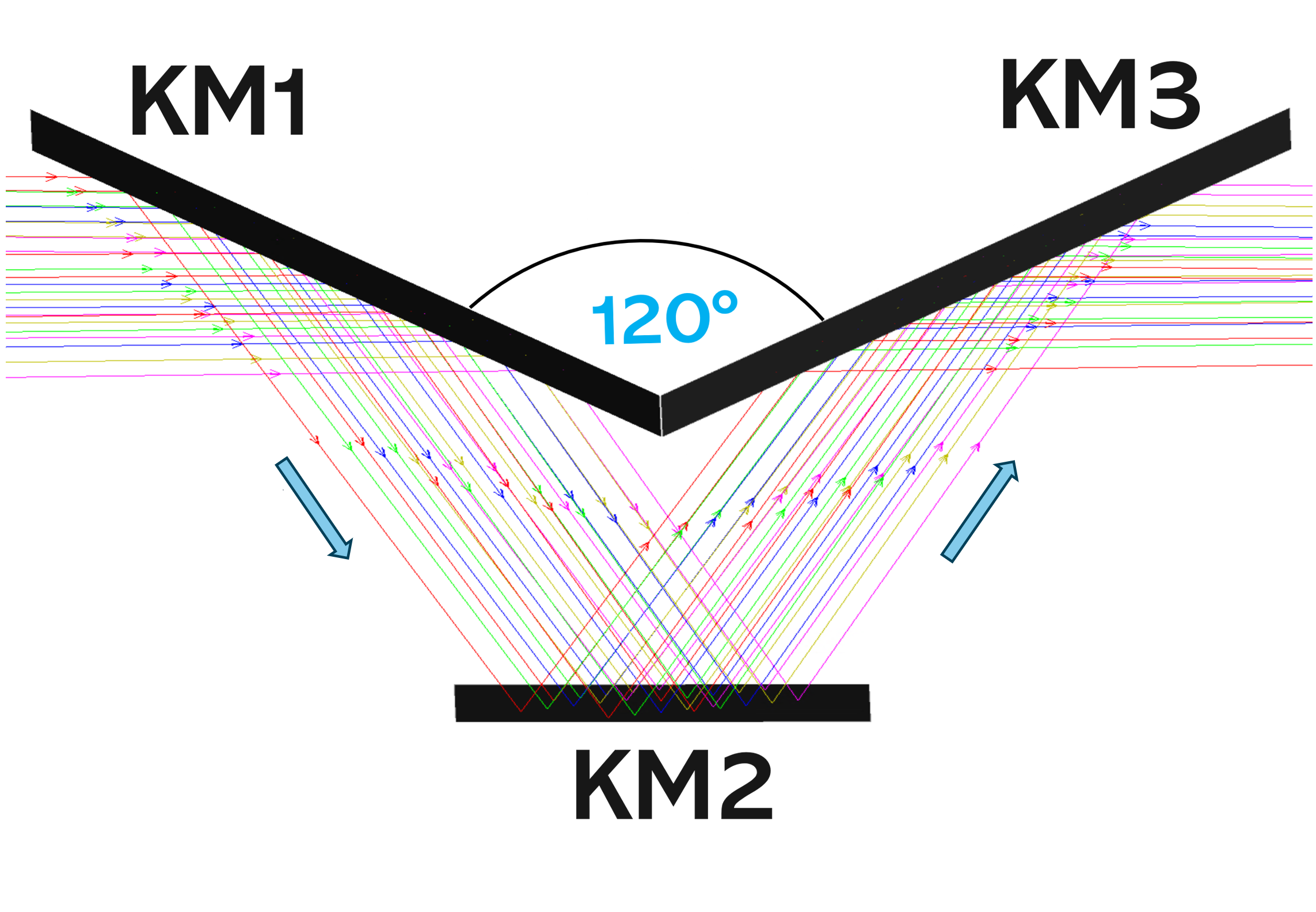}
\caption{\footnotesize Optical layout of the reflective three-mirror K-mirror image rotator.}
\label{fig:kmirror}
\vspace{-15pt}
\end{wrapfigure}
Image rotation in astronomical instruments is commonly achieved using either refractive elements, such as Dove prisms, or reflective systems, such as K-mirror assemblies \cite{Karan_2022}. Although a Dove-prism design was initially considered for the Tull Spectrograph, the broad operational wavelength range (350--1000\,nm) and the desire to avoid additional refractive elements motivated the adoption of a reflective approach. A K-mirror configuration was therefore investigated as a potential future upgrade to maintain a fixed slit orientation for observations of extended targets. Using only reflective surfaces, the K-mirror introduces no chromatic aberration, preserves broadband throughput, and provides continuous image rotation while remaining compatible with the existing coud\'e optical path.

\noindent The image rotates at the HJST coude focus, and so image derotation is important for extended targets (eg. comets and planets) and targets in confused fields (eg. stars in globular clusters) to preserve consistent target alignment and stable slit illumination during long integrations. HJST coude has had a K-mirror image rotator since 1969, but we now need a computerized
 image rotator with modern high reflectivity coatings. To address these requirements, a reflective three-mirror K-mirror image rotator was designed for integration within the HJST coud\'e optical relay. The optical layout of the K-mirror assembly is shown in Figure ~\ref{fig:kmirror}. The system consists of two identical outer fold mirrors (KM1 and KM3), each have dimensions of 70 mm × 33 mm and a central folding mirror (KM2) which has dimensions of 40 mm × 30 mm, arranged in a symmetric folded geometry. The beam undergoes three successive reflections before exiting parallel to the incident beam, enabling image derotation while preserving the nominal f/32.32 coud\'e beam propagation. The extra optical path length is easily accomodated by a change in telescope
focus which can be made without any image quality degradation.

\noindent The image rotation introduced by the K-mirror is related to the mechanical rotation angle of the assembly through
$\theta_{\mathrm{image}} = 2\theta_{\mathrm{mechanical}}$
allowing continuous field derotation during telescope tracking.  The geometry of the K-mirror was determined based on the required beam clearance, field of view, and available \begin{wrapfigure}{r}{0.30\textwidth}
\vspace{-1pt}
\centering
\includegraphics[width=0.7\linewidth]{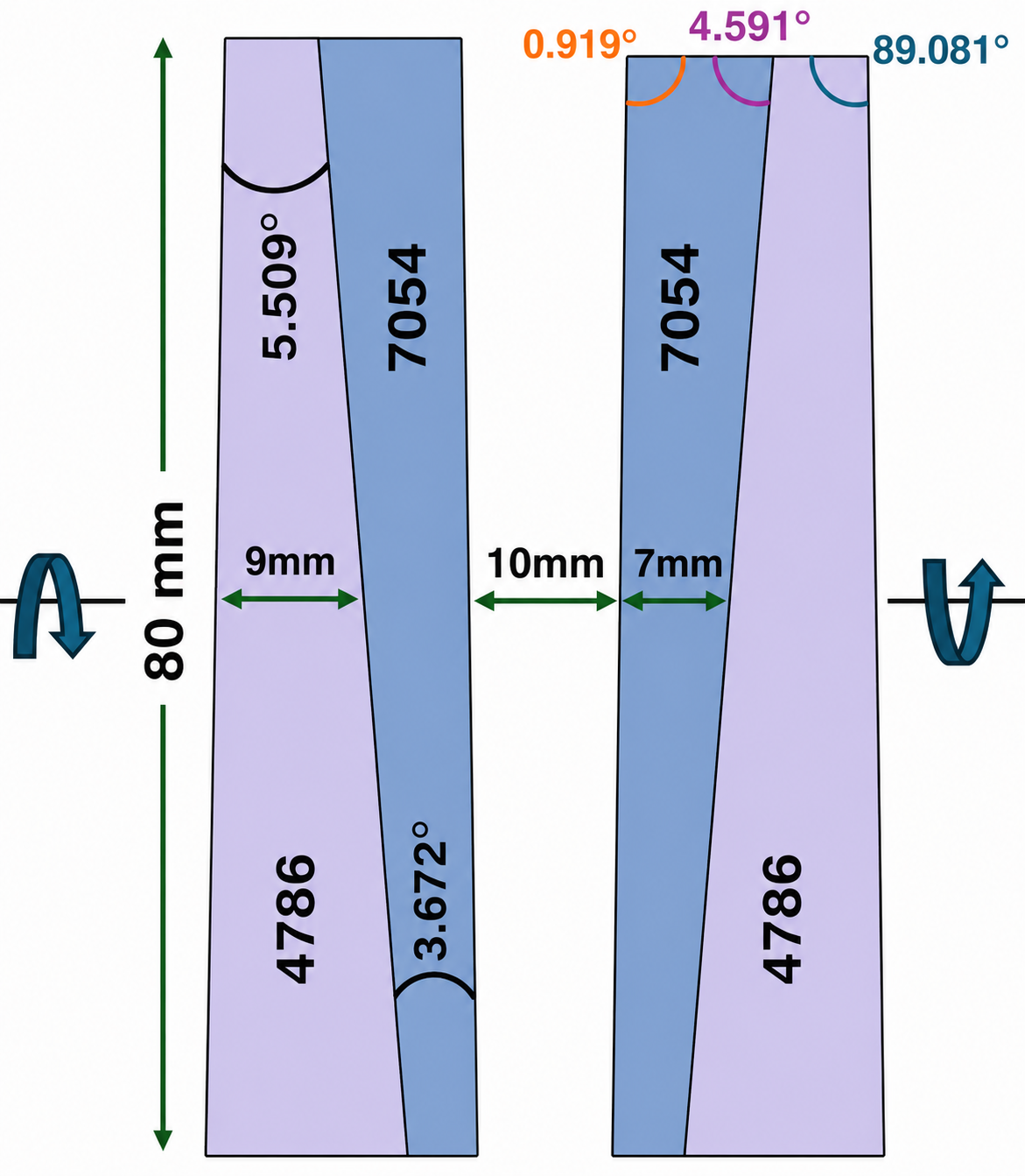}
\caption{\footnotesize
Conceptual layout of the RADC consisting of two identical counter-rotating Amici prism assemblies.
}
\label{fig:radc_layout}
\vspace{-15pt}
\end{wrapfigure} mechanical envelope within the existing coud\'e relay. The propagation of the beam through the f/32.32 relay was evaluated to determine the minimum clear apertures required to avoid vignetting across the operational field.
The design supports an approximate field of view of 1 arcmin and is optimized to maintain stable slit illumination and target alignment during long-exposure high-resolution spectroscopic observations, regardless of telescope hour angle. Such image-orientation control is particularly beneficial for observations of extended targets, including stellar clusters, nearby galaxies, and nebular objects, where a fixed slit position angle enables consistent sampling of selected regions within the field throughout an exposure. From a mechanical integration perspective, a packaging study determined that the K-mirror assembly can be accommodated within the existing coud\'e  relay between the calibration subsystem and the proposed ADC assembly. The available envelope provides approximately 30\,mm of clearance relative to the ADC package, and the existing alignment-laser mounting location is being investigated as a potential installation point. 

\subsection{ROTATIONAL ATMOSPHERIC DISPERSION CORRECTOR}

\begin{wraptable}{l}{0.50\textwidth}
\vspace{-13pt}
\caption{\footnotesize Key specifications of the RADC designed for the Tull Spectrograph.}
\label{tab:radc_specs}
\footnotesize
\begin{tabular}{|c|c|}
\hline
\textbf{Parameter} & \textbf{Specification} \\
\hline
Zenith angle correction range & 0$^\circ$ - 65.5$^\circ$ \\
Field of view & 2 arcmin \\
Lens diameter & 80 mm \\
Prism materials & Nikon 4786 \& Nikon 7054 \\
Central air gap & 10 mm \\
Prism central thickness & 9 mm / 7 mm \\
Prism apex angles & 5.509$^\circ$, 3.672$^\circ$ \\
Prism wedge angles & 0.919$^\circ$, 4.591$^\circ$ \\
\hline
\end{tabular}
\vspace{-10pt}
\end{wraptable}
Atmospheric dispersion in astronomical instruments is commonly corrected using either Longitudinal Atmospheric Dispersion Correctors (LADCs) \cite{Avila_1997,Phillips_2006} or Rotational Atmospheric Dispersion Correctors (RADCs) \cite{Wynne_1993,Bestha_2025}. LADCs compensate dispersion by varying the separation between oppositely oriented prisms in a converging beam, but at the Tull Spectrograph's f/32 beam they require substantial axial travel, increasing instrument length and introducing beam displacement that must be compensated by the telescope control system. Given the limited space available within the existing coudé optical path and the requirement to preserve the nominal beam geometry, a Rotational Atmospheric Dispersion Corrector (RADC) was selected, providing continuous dispersion correction with compact packaging and negligible beam displacement. The RADC is designed to compensate wavelength-dependent atmospheric refraction in the HJST coud\'e beam prior to injection into the Tull Spectrograph slit. Atmospheric dispersion increases with zenith angle and produces chromatic elongation of the stellar image, resulting in wavelength-dependent slit losses and reduced throughput in high-resolution spectroscopy. Across the operational wavelength range of 350--1000\,nm, the uncorrected atmospheric dispersion can exceed the nominal slit width at large zenith angles, making broadband chromatic correction essential for maintaining stable slit illumination and spectral fidelity. The proposed RADC employs two identical counter-rotating Amici prism assemblies to provide continuous atmospheric-dispersion correction over the required zenith range of 0$^\circ$--65.5$^\circ$. By varying the relative prism rotation angles, the system produces adjustable chromatic deviation that compensates atmospheric refraction while preserving the nominal f/32.32 coud\'e beam geometry. The refractive configuration provides an unvignetted field of view of 2\,arcmin and is integrated within the existing coud\'e optical path. The final design parameters are summarized in Table~\ref{tab:radc_specs}.

\begin{figure*}[t]
\centering
\includegraphics[width=0.8\textwidth]{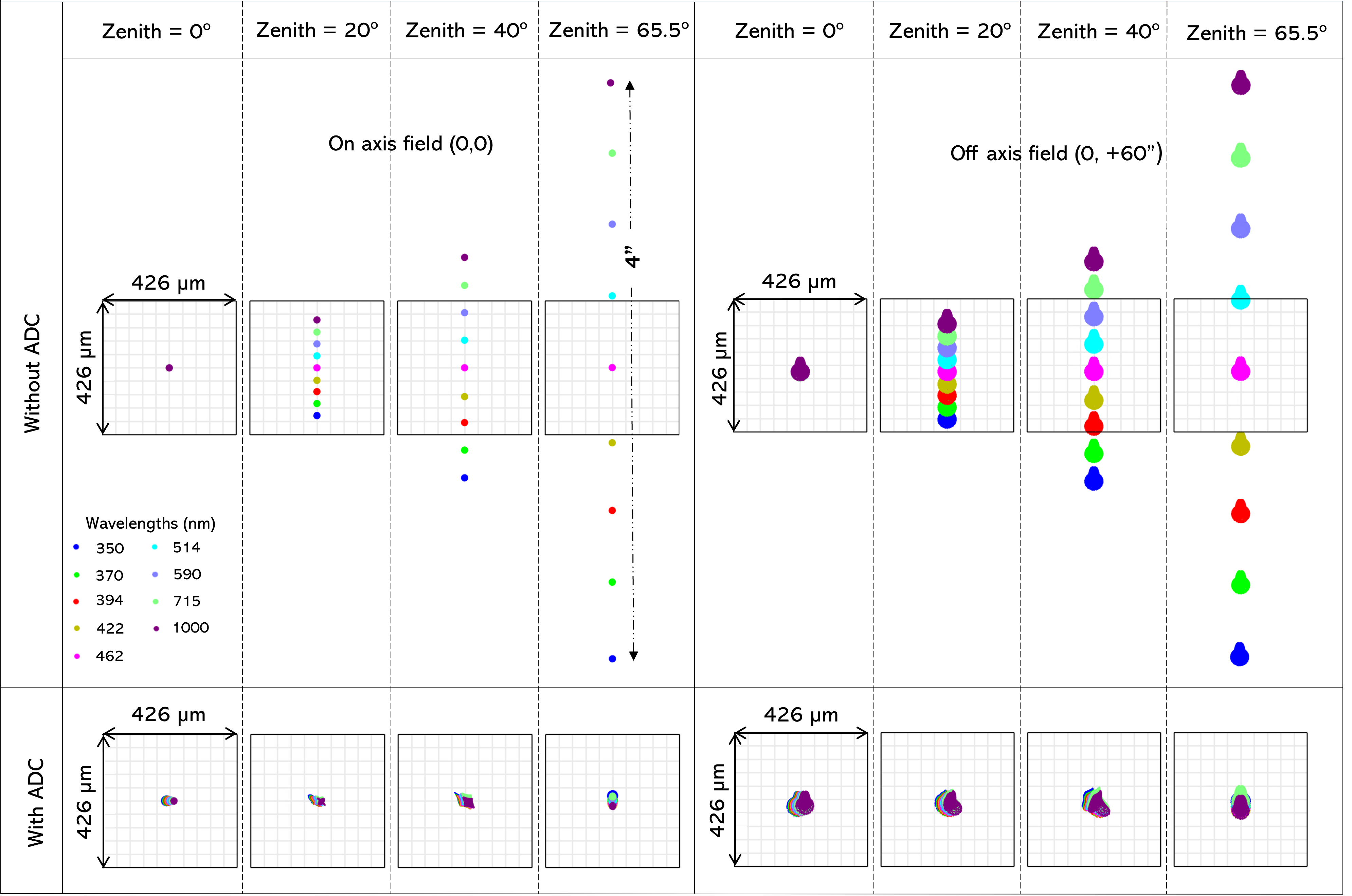}
\caption{\footnotesize  Polychromatic image quality of the HJST RADC evaluated at the maximum design zenith angle of $65.5^\circ$. The left and right panels correspond to on-axis $(0,0)$ and off-axis $(0,+60'')$ field positions, respectively. The small off-axis coma is primarily due to the telescope not being Ritchey-Cretein at its coude focus. In the upper panels, colored markers represent image centroid locations over the wavelength range 350--1000 nm after atmospheric-dispersion correction. For reference, a 1 arcsec (426\,$\textmu $m) scale is shown and the approximate magnitude of the uncorrected atmospheric dispersion ($\sim4^{\prime\prime}$) are indicated. The lower panels show the corresponding residual spot distributions after correction. The ADC maintains image quality while reducing atmospheric dispersion to levels that are small compared with the R $\sim$ 60,000 slit width of 1.18$^{\prime\prime}$ and typical seeing conditions (1.3$^{\prime\prime}$).
}
\label{fig:adc_image_quality}
\vspace{-5pt}
\end{figure*}
\section{PERFORMANCE ANALYSIS}
\label{sec:analysis}

\textbf{Image Quality}: The image quality performance of the ADC was evaluated using polychromatic Zemax OpticStudio spot diagrams over the wavelength range 350--1000\,nm for both on-axis and off-axis field positions within the corrected field of view. Figure~\ref{fig:adc_image_quality} compares the wavelength-dependent image displacement before correction with the residual spot distributions after atmospheric-dispersion correction for zenith angles ranging from $0^\circ$ to the maximum design value of $65.5^\circ$. The upper row illustrates the chromatic image displacement produced by atmospheric refraction in the absence of correction. As expected, the magnitude of atmospheric dispersion increases with zenith angle, reaching approximately 4\,arcsec at $65.5^\circ$. For reference, a 1 arcsec (426 um) scale is shown. The lower row shows the corresponding residual spot diagrams after application of the optimized ADC for various zenith angles and field positions considered. The corrected spots exhibit only small residual chromatic offsets and negligible degradation of image quality across the field of view. 

\noindent \textbf{Dispersion Correction \& Geometric Throughput Loss}: 
The variation of parallactic angle during telescope tracking results in a continuous rotation of the atmospheric-dispersion vector relative to the spectrograph\cite{Filippenko_1982}. For broadband observations, this changing dispersion orientation can introduce wavelength-dependent slit losses and non-uniform slit illumination, particularly at large zenith angles.

\noindent The parallactic angle\footnote{\url{https://en.wikipedia.org/wiki/Parallactic_angle}} $q$ was calculated using

\begin{equation}
\tan q =
\frac{\sin H}
{\tan \phi \cos \delta - \sin \delta \cos H}
\label{eq:parallactic_angle}
\end{equation}

\noindent where $H$ is the hour angle, $\phi$ is the observatory latitude, and $\delta$ is the target declination. For the HJST at McDonald Observatory, the observatory latitude was taken as $\phi = 30.67^\circ$.
\begin{figure}[ht]
\centering
\includegraphics[width=0.40\linewidth]{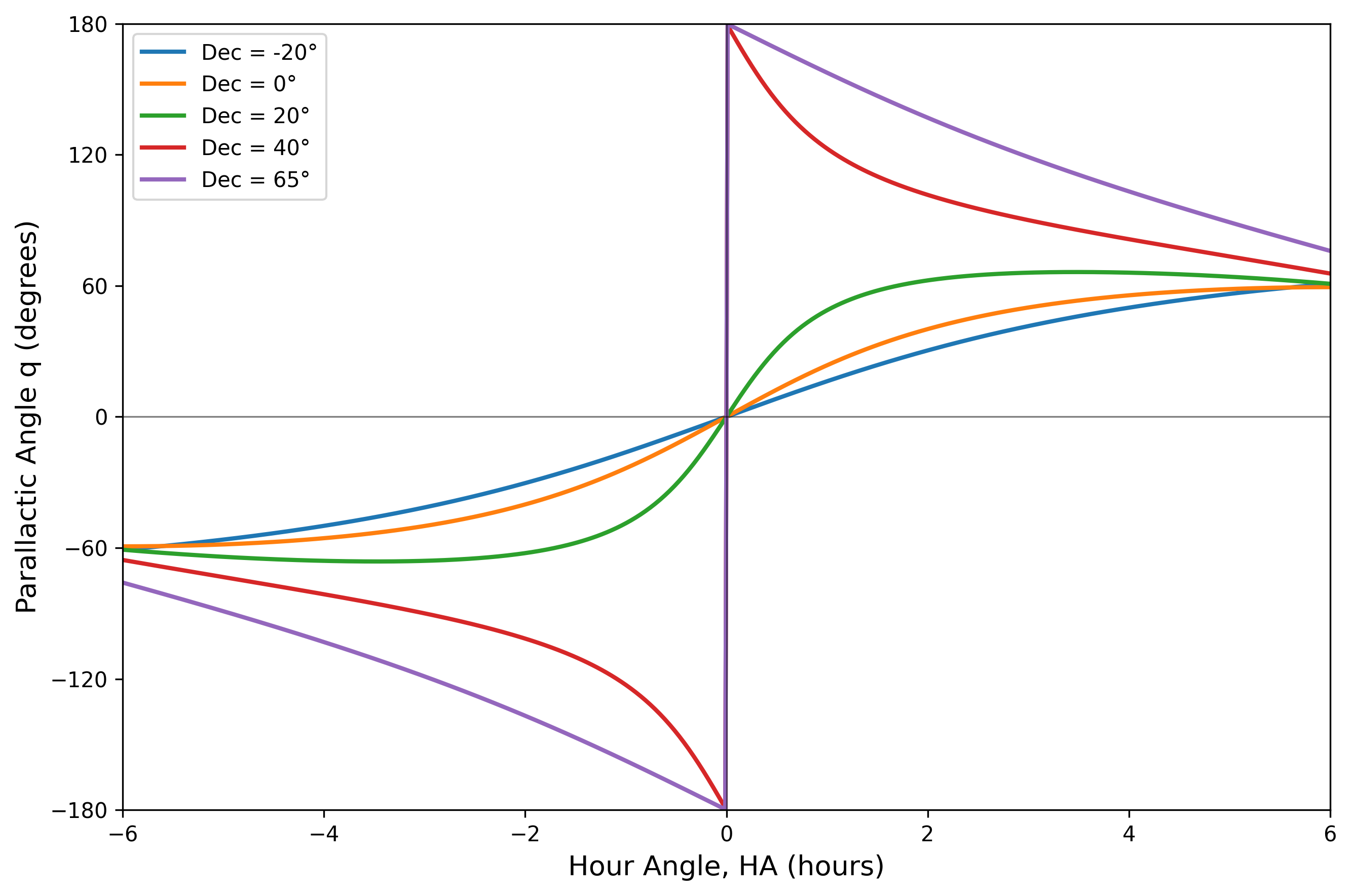}
\includegraphics[width=0.40\linewidth]{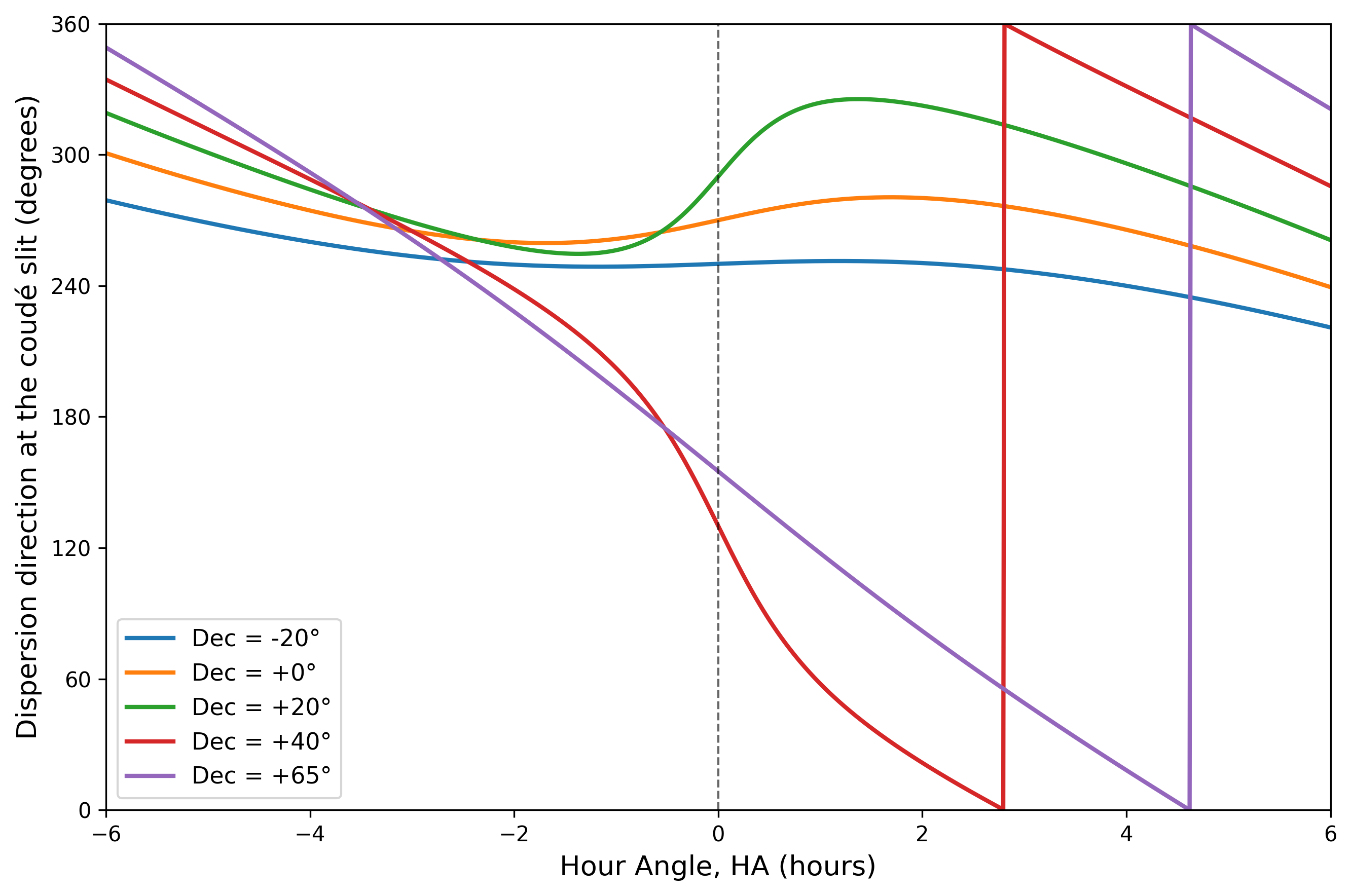}
\caption{\footnotesize
(\textit{Left}) Classical parallactic angle $q$, computed using Equation~\ref{eq:parallactic_angle}, as a function of hour angle for targets with declinations ranging from $-20^\circ$ to $+65^\circ$ at the Mount Locke latitude of $\phi = 30.67^\circ$. (\textit{Right}) Corresponding atmospheric-dispersion orientation ptojected onto the HJST coud\'e slit plane using the adopted telescope position-angle convention of 0 degree at the top of the slit.}
\label{fig:parallactic_angle}
\end{figure}
\begin{figure*}[ht]
\centering
\includegraphics[width=0.75\linewidth]{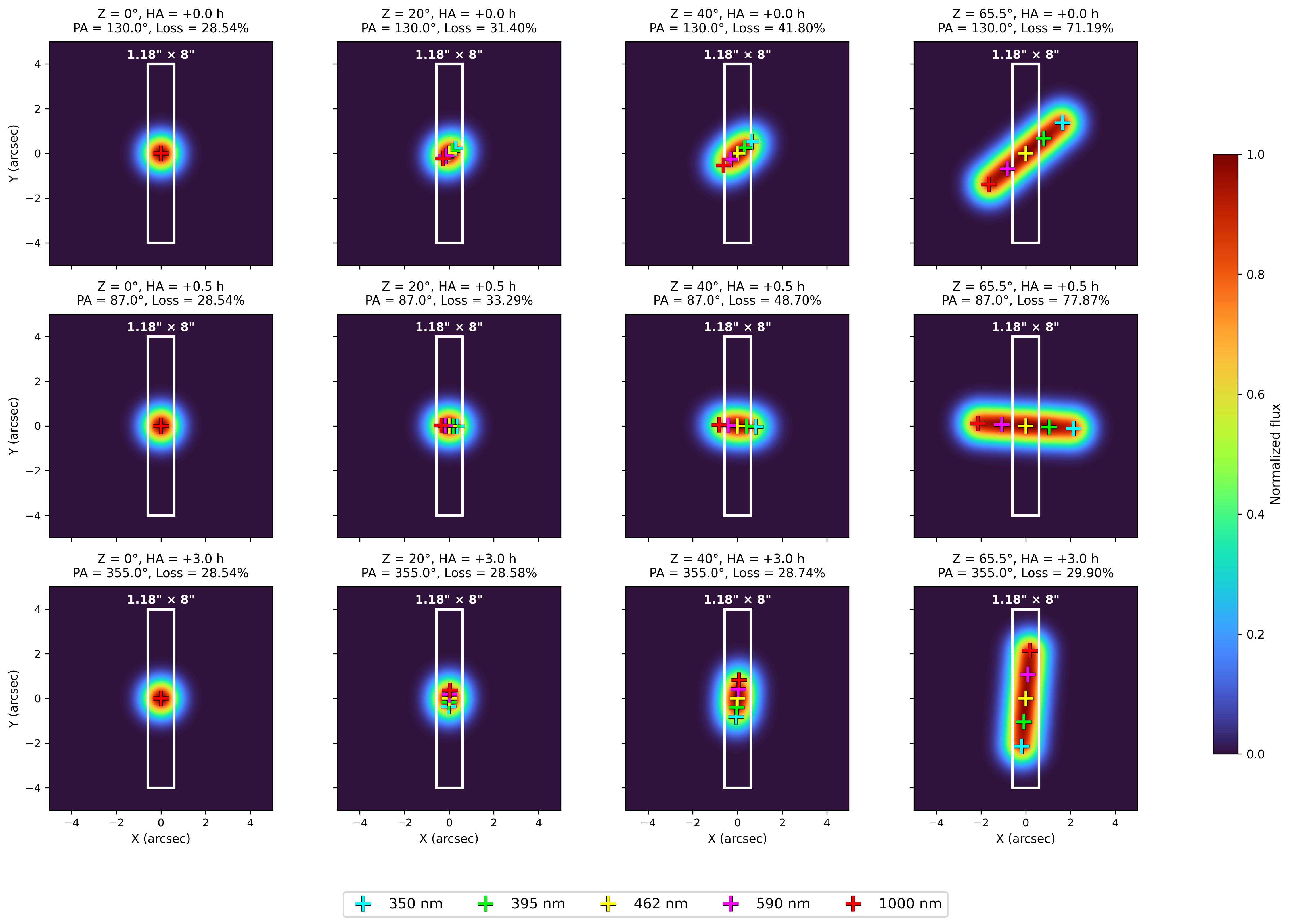}
\caption{\footnotesize
Simulated wavelength-dependent slit losses for the Tull Spectrograph as a function of zenith angle and hour angle for a target declination of $+40^\circ$. Each panel shows the seeing-convolved combined point spread function for a $1.8^{\prime\prime} \times 8^{\prime\prime}$ slit geometry, with colored ``+'' symbols indicate the image centroid positions at the sampled wavelengths of 350, 395, 462, 590, and 1000 nm. A typical seeing FWHM of 1.3$''$ was assumed. Rows correspond to different hour angles, illustrating the changing orientation of atmospheric dispersion during telescope tracking, while columns represent increasing zenith angle. At large zenith distances, the atmospheric dispersion becomes increasingly extended and rotated relative to the slit geometry, producing significant wavelength-dependent throughput losses when the dispersion direction approaches the slit-width axis.
}
\vspace{-10pt}
\label{fig:dispersion_uncorrected}
\end{figure*}
\noindent Figure~\ref{fig:parallactic_angle} shows the variation of parallactic angle with hour angle for various target declinations. The orientation of atmospheric dispersion changes continuously during tracking and varies significantly with target declination. Near meridian transit, high-declination targets exhibit rapid changes in parallactic angle, producing substantial rotation of the atmospheric dispersion direction relative to the slit orientation. The orientation of atmospheric dispersion relative to the fixed coud\'e slit was evaluated by combining the classical parallactic-angle formalism with the telescope position-angle convention adopted for the HJST, as illustrated in the right-hand panel of Figure~\ref{fig:parallactic_angle}. While the parallactic angle defines the dispersion direction in the celestial reference frame, the projected orientation at the coud\'e slit differs because of the coordinate transformation introduced by the telescope mounting geometry and coud\'e optical path.

\begin{figure*}[t]
\centering
\includegraphics[width=0.75\textwidth]{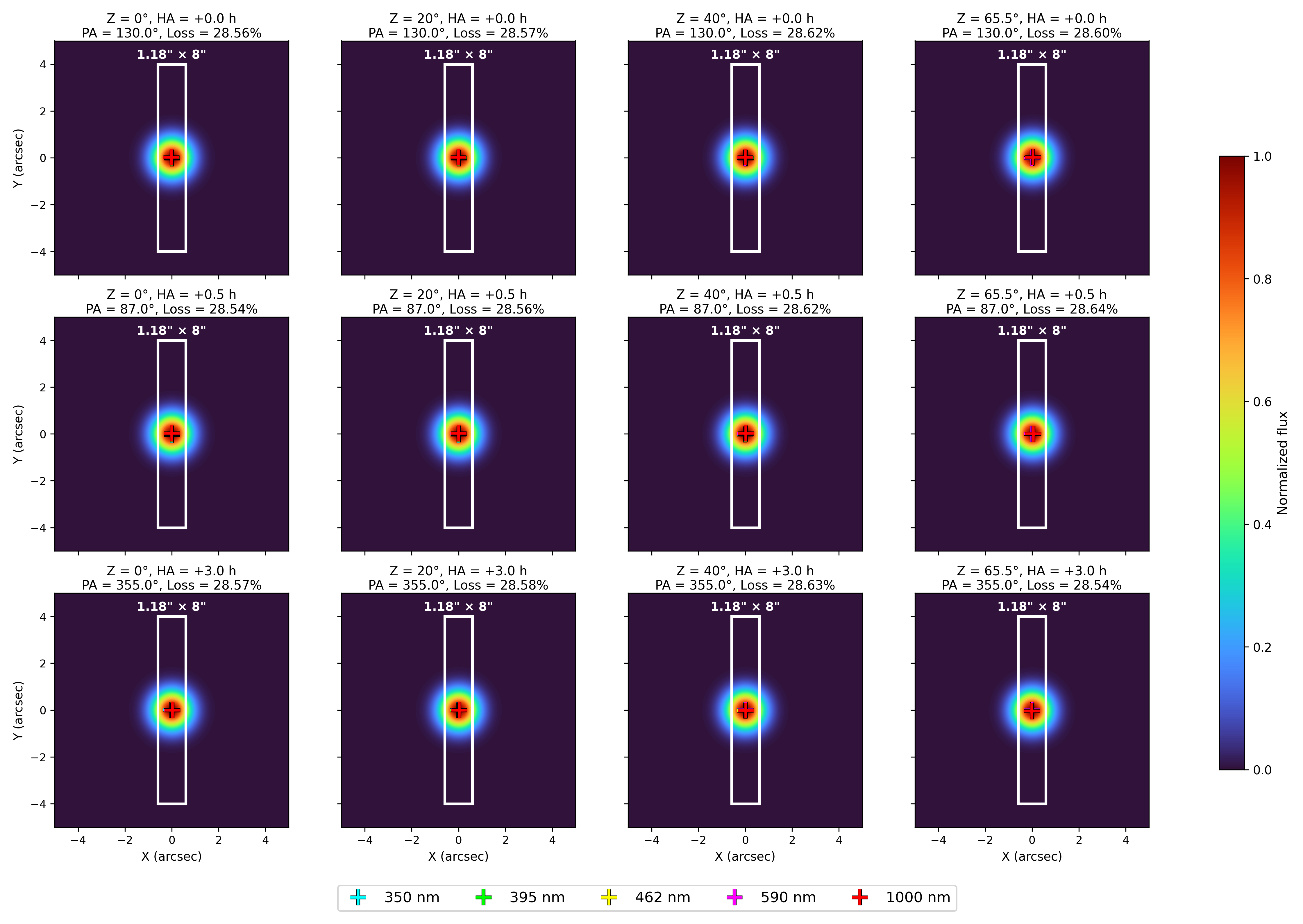}
\caption{\footnotesize Simulated seeing-convolved slit illumination after atmospheric-dispersion correction for the Tull Spectrograph using a $1.18^{\prime\prime} \times 8^{\prime\prime}$ slit geometry. The compensated point spread functions remain well confined within the slit aperture across all observing geometries, indicating effective suppression of atmospheric dispersion. As a result, the slit losses remain nearly constant at $\sim$28.5\%--28.6\%.}
\label{fig:adc_compensated}
\vspace{-15pt}
\end{figure*}

\noindent Figure~\ref{fig:dispersion_uncorrected} shows the effect of atmospheric dispersion on slit throughput for the Tull Spectrograph. The analysis employed Zemax OpticStudio spot centroids combined with seeing-convolved point-spread-function simulations assuming a typical seeing FWHM of 1.3 arcsec.  The spot centroids were extracted from the OpticStudio model using the PyZDDE Python interface\footnote{\url{https://pypi.org/project/PyZDDE/}}. A slit geometry of $1.18^{\prime\prime}\times8^{\prime\prime}$, corresponding to the TSF3 slit used for $R\sim60,000$, was adopted. The simulations demonstrate that the changing parallactic geometry during telescope tracking can introduce significant chromatic throughput variations. These effects become increasingly severe at large zenith distances and motivate the implementation of an atmospheric dispersion correction strategy to maintain stable slit illumination and minimize wavelength-dependent throughput losses during long spectroscopic integrations.

\noindent The effectiveness of the ADC was evaluated through seeing-convolved slit-throughput simulations using Zemax OpticStudio-derived wavelength centroid positions projected onto the HJST coud\'e slit plane as shown in Figure \ref{fig:adc_compensated}. 
In contrast to the uncompensated case shown in Figure~\ref{fig:dispersion_uncorrected}, the compensated simulations demonstrate that the wavelength centroids remain nearly co-spatial within the slit aperture even at large zenith angles, indicating substantial reduction of chromatic differential refraction at the coud\'e slit plane. 
The calculated light loss at the slit remains nearly constant at $\sim$28.5\%, compared to the significantly larger and strongly geometry-dependent (geometry and wavelength dependent) losses obtained without atmospheric-dispersion correction ($>$ $70\%$). 
\begin{figure*}[t!]
\centering
\includegraphics[width=0.31\textwidth]{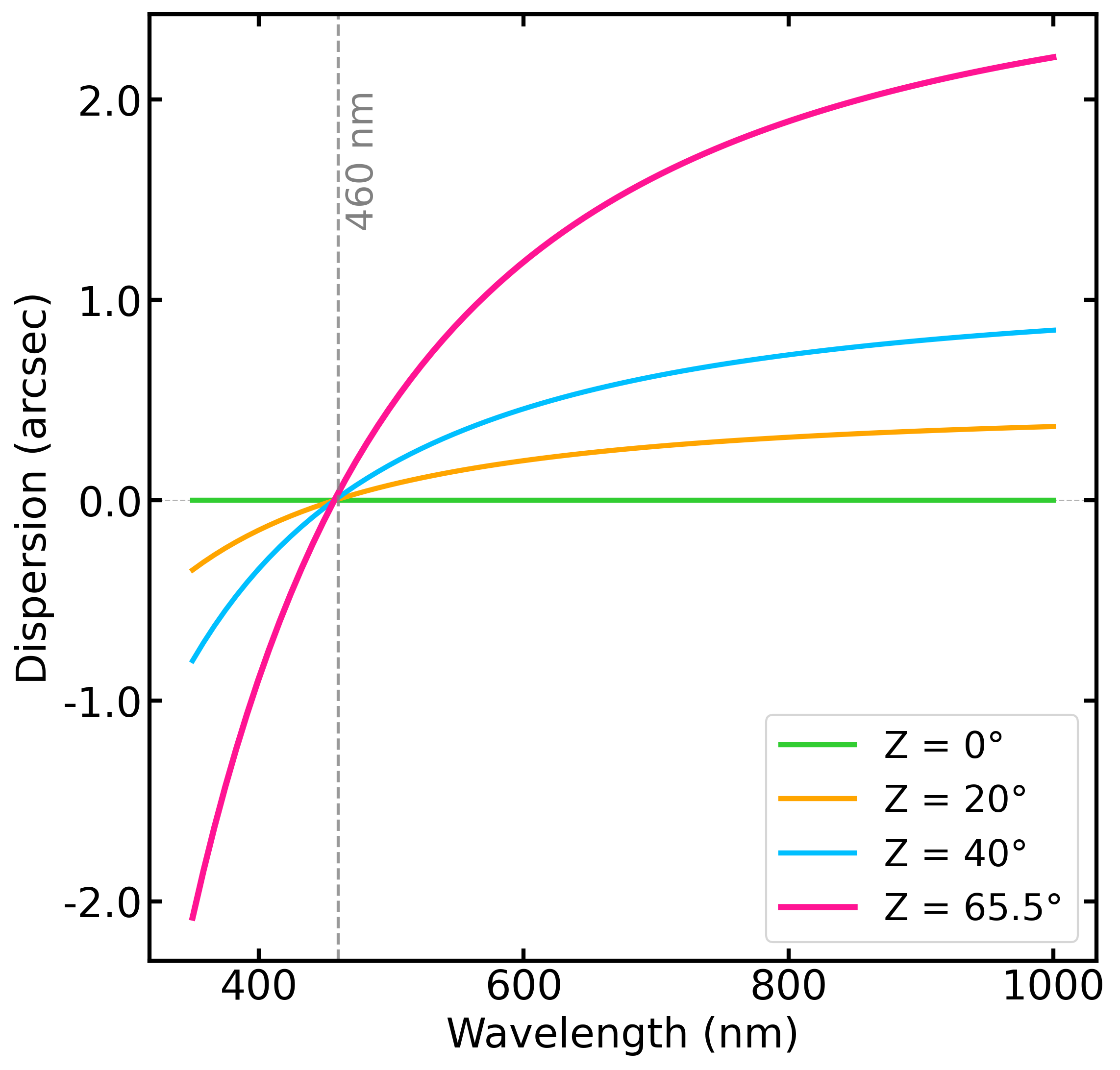}
\includegraphics[width=0.315\textwidth]{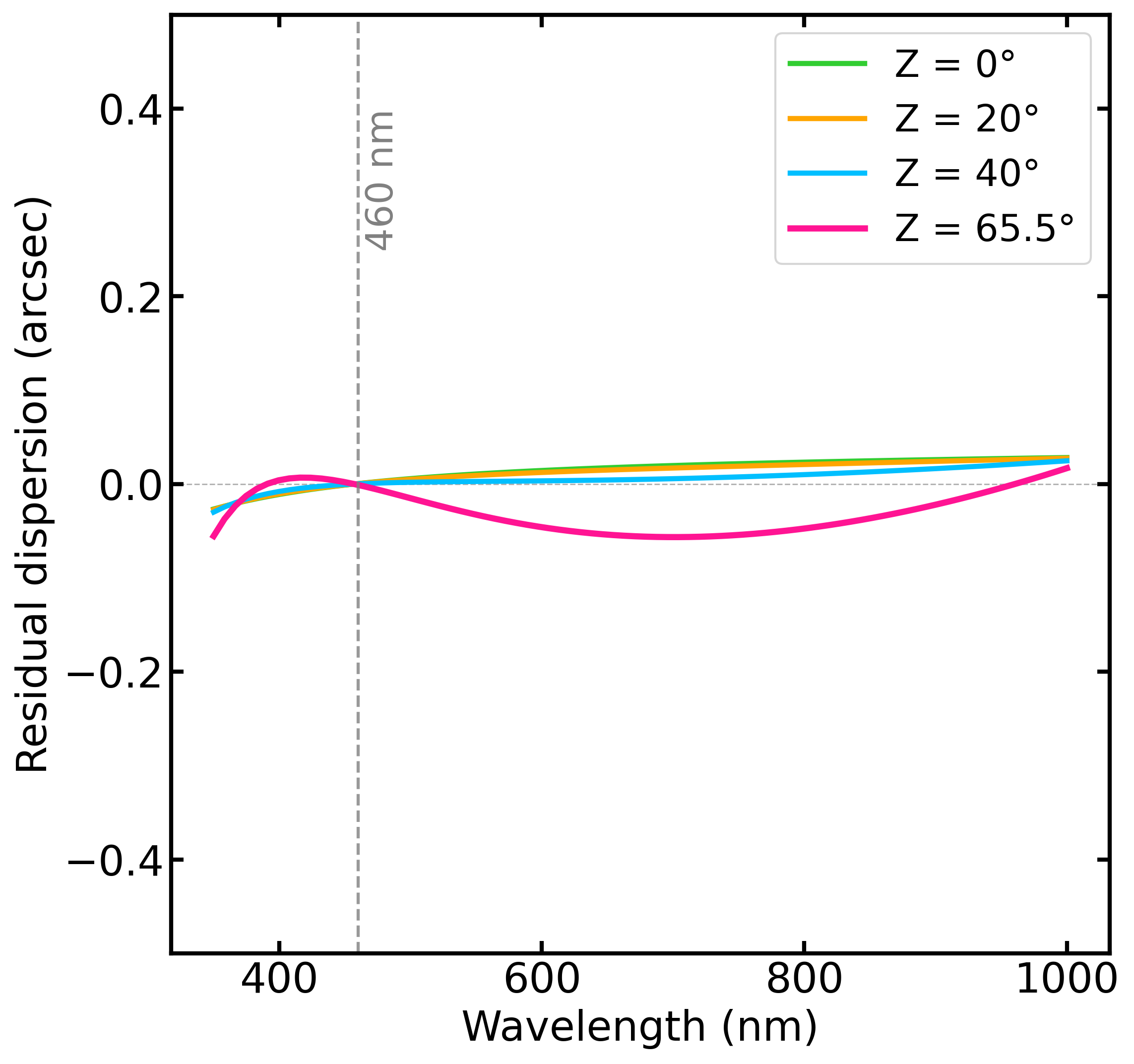}
\includegraphics[width=0.31\textwidth]{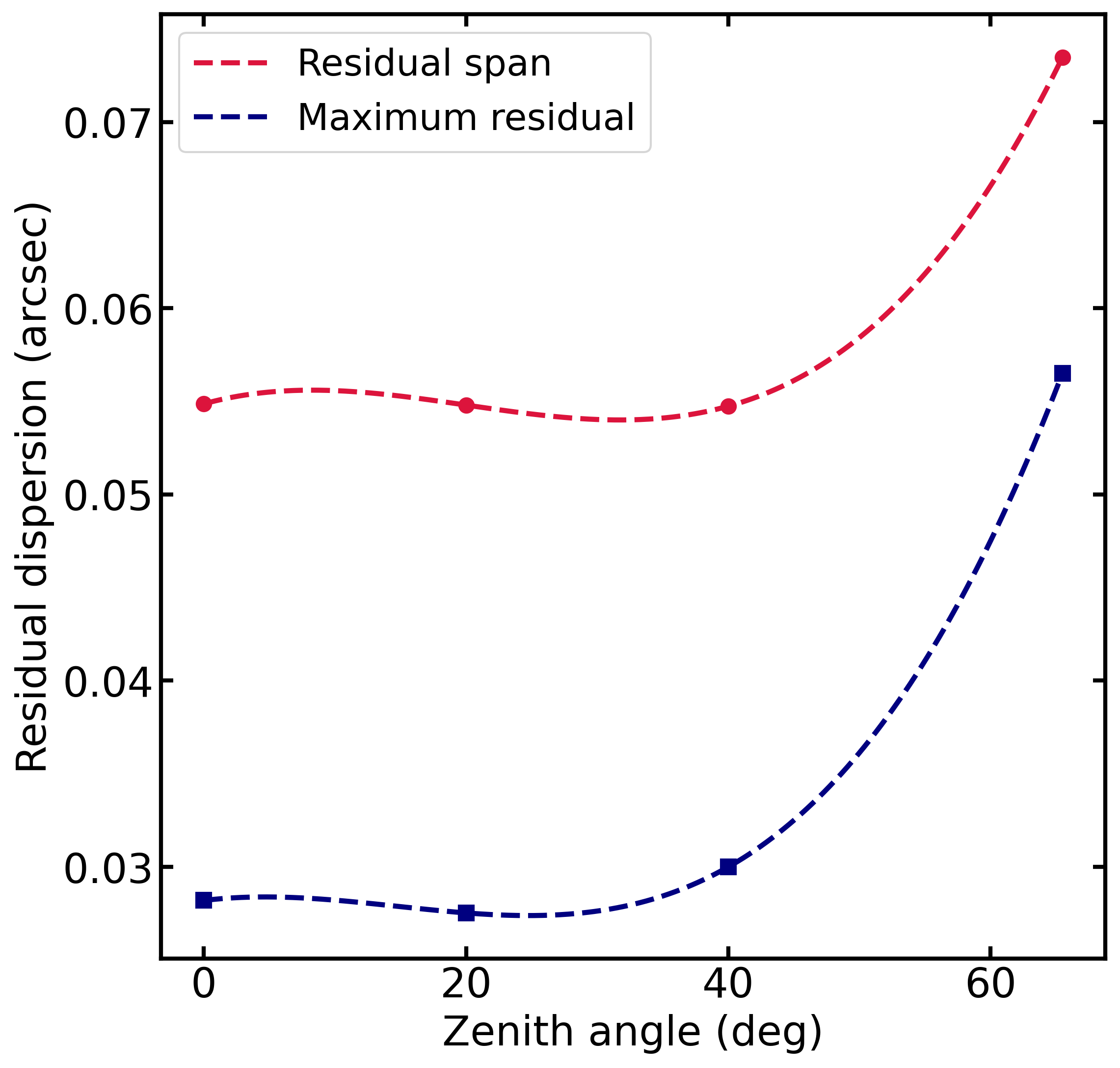}
\caption{\footnotesize
Atmospheric dispersion correction performance of the HJST ADC. (a) Uncorrected atmospheric dispersion as a function of wavelength for representative zenith angles. (b) Residual chromatic displacement after ADC correction, referenced to the optimization wavelength of 460\,nm. (c) Residual dispersion metrics as a function of zenith angle, including the residual span across the full wavelength range and the maximum residual displacement relative to the reference wavelength.
}
\label{fig:adc_performance}
\vspace{-15pt}
\end{figure*}

\noindent Figure~\ref{fig:adc_performance} summarizes the atmospheric dispersion correction performance of the HJST ADC over the wavelength range 350--1000 nm. Prior to correction, atmospheric dispersion increases rapidly with zenith angle, reaching a total chromatic displacement of approximately 4.3\,arcsec at the maximum design zenith angle of $65.5^\circ$. Such dispersion significantly exceeds the slit width and would result in substantial wavelength-dependent slit losses. Following optimization of the ADC, the residual chromatic displacement is reduced by more than two orders of magnitude across the full wavelength range. At all design zenith angles, the residual dispersion remains below $\pm0.06$\,arcsec relative to the reference wavelength of 460 nm. The residual dispersion span (dispersion between extreme blue and red wavelengths) remains approximately 0.055\,arcsec over most of the operating range and increases slightly to approximately 0.074\,arcsec at the largest zenith angle. Similarly, the maximum residual dispersion relative to the reference wavelength 460 nm remains below approximately 0.057\,arcsec across the full operating range. These results demonstrate that the ADC effectively suppresses atmospheric dispersion to levels that are small compared with both the slit width and typical atmospheric seeing, thereby preserving throughput and minimizing chromatic slit losses.\\
\begin{wrapfigure}{r}{0.34\textwidth}
\vspace{-10pt}
\centering
\includegraphics[width=0.32\textwidth]{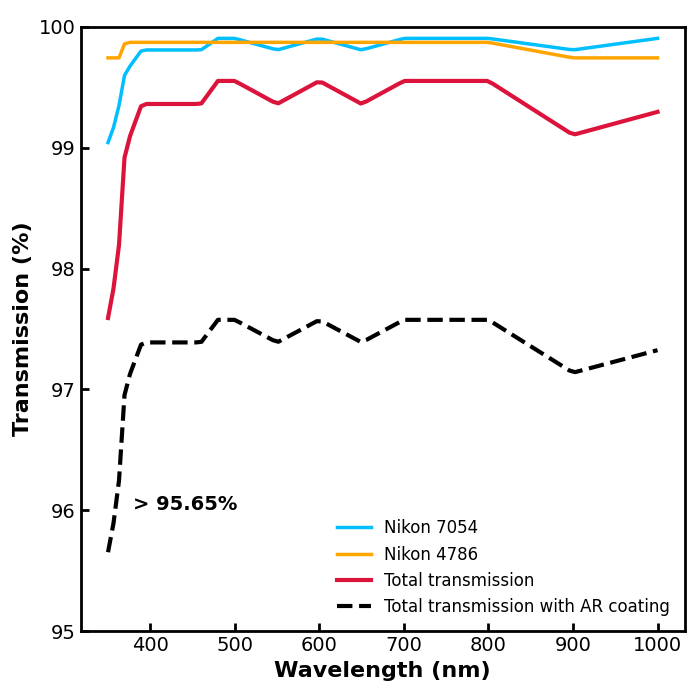}
\caption{\footnotesize
Predicted HJST RADC transmission across 350--1000\,nm, remaining above 95.65\% even after accounting for coating losses.
}
\label{fig:adc_transmission}
\vspace{-15pt}
\end{wrapfigure}
\textbf{Transmission Curve}: The transmission performance of the RADC was evaluated over the operational wavelength range of 350--1000 nm using the internal transmission characteristics of the selected prism materials, Nikon 4786 and Nikon 7054. 

\noindent Figure~\ref{fig:adc_transmission} presents the transmission of the individual prism materials together with the predicted transmission of the complete ADC assembly. The blue and orange curves show the internal transmission of the Nikon 7054 and Nikon 4786 prism materials, respectively. Both optical glasses exhibit excellent broadband transmission, exceeding approximately 99.8\% over most of the wavelength range (390 - 900 nm). The combined transmission of the ADC materials remains above approximately 99.1\% over the full 350--1000\,nm wavelength range. The red curve represents the combined transmission of the complete ADC assembly, while the dashed black curve includes estimated losses from broadband anti-reflection coatings on all optical surfaces. High transmission in the near-UV and blue wavelength region was a primary criterion in the glass-selection process, leading to the adoption of i-line optical glasses\footnote{\url{https://www.nikon.com/business/components/lineup/materials/i-line/}}. To estimate the end-to-end throughput of the system, broadband anti-reflection (AR) coating losses were included for all optical surfaces. A conservative coating loss of 0.5\% per surface was assumed in the analysis\footnote{\url{https://spectrumthinfilms.com/stf/astronomical-coatings-optics/}}. After accounting for coating losses, the transmission of the complete ADC assembly remains above 95.65\% across the full wavelength range. 

\section{MECHANICAL DESIGN}
\begin{figure*}[t]
\centering
\includegraphics[width=0.65\textwidth]{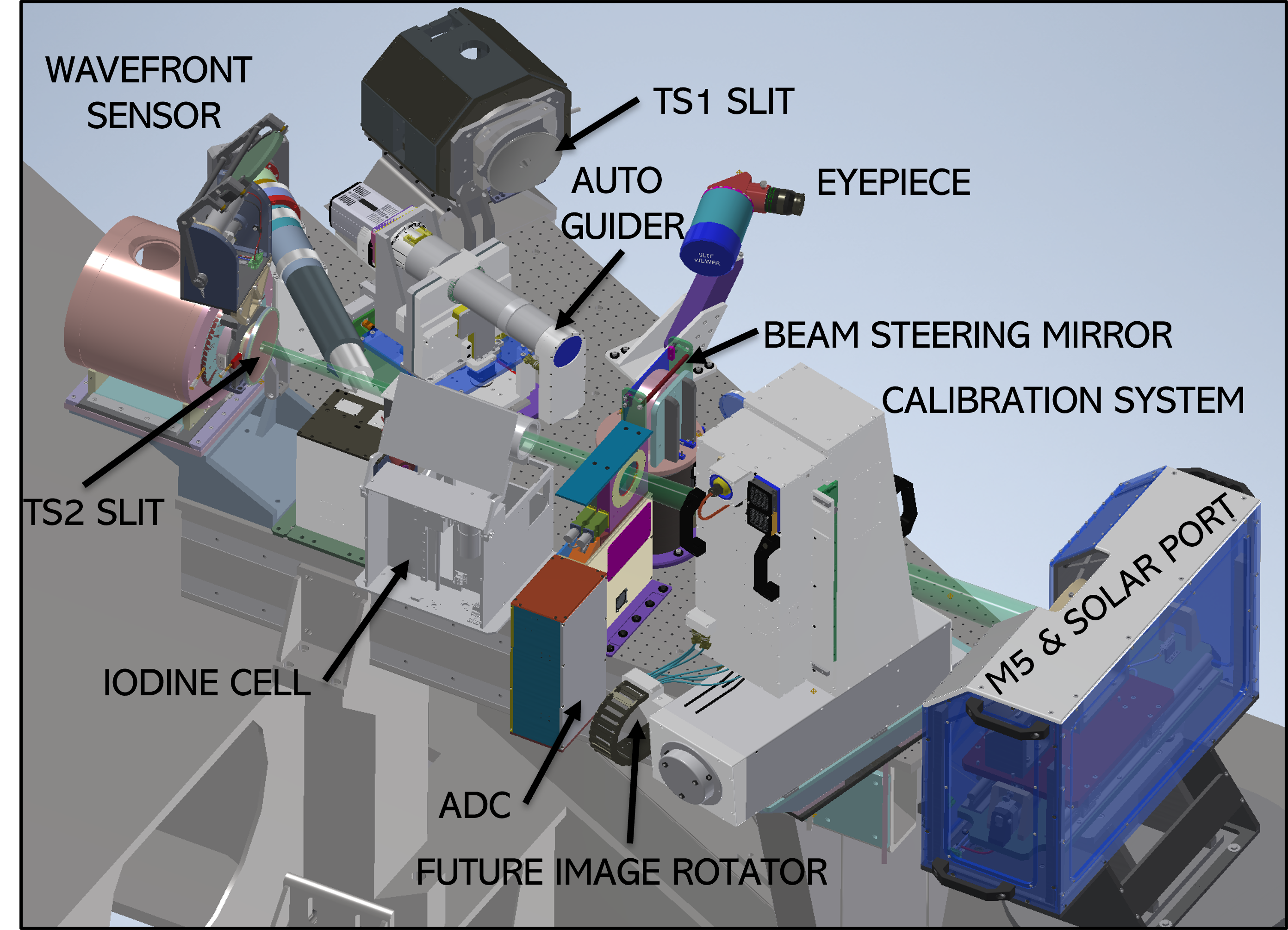}
\caption{\footnotesize
Opto-mechanical layout of the HJST coud\'e spectrograph slit bench showing the major subsystems integrated within the coud\'e relay. The telescope beam enters from the M5 and solar-port interface and passes through the deployable Atmospheric Dispersion Corrector (ADC). The slit bench incorporates dual observing modes through the TS1 (retired) and TS slit assemblies, an auto-guider, wavefront sensor, calibration injection system, iodine cell module, and visual eyepiece. The ADC is positioned upstream of the slit to compensate atmospheric dispersion prior to target acquisition and spectroscopy while preserving compatibility with existing optics and calibration subsystems.
}
\label{fig:slitbench_layout}
\end{figure*}

The atmospheric dispersion corrector is being integrated within the existing Tull Spectrograph preslit optics. Figure~\ref{fig:slitbench_layout} illustrates the current optical bench configuration together with the location of the ADC within the relay. Light from the telescope enters the coud\'e slit room, reflects off M5 and propagates through the existing calibration and acquisition subsystems before reaching the spectrograph slit. The optical bench includes several operational subsystems, including the calibration unit, iodine cell, wavefront sensor, guider beam-steering mirror, autoguider, eyepiece assembly, and the TS and former TS1 slit stations. The ADC is positioned between the M5 mirror and the downstream slit assemblies, enabling atmospheric-dispersion correction to be applied before slit injection. This location was selected to maximize the effectiveness of the correction while preserving compatibility with the existing optical layout and avoiding interference with the guider and calibration systems.
Integration within the available mechanical envelope of the coud\'e relay was a key design requirement, ensuring that existing observing modes and auxiliary subsystems remain fully operational. The available mechanical envelope within the relay constrained the overall package size and placement of the correction optics. To preserve the existing beam geometry and minimize vignetting risk, the assembly was designed with minimal protrusion beyond the final optical surface toward the slit to reserve space for a future fast tip-tilt system.

\begin{figure}[t]
\centering
\includegraphics[width=0.58\textwidth]{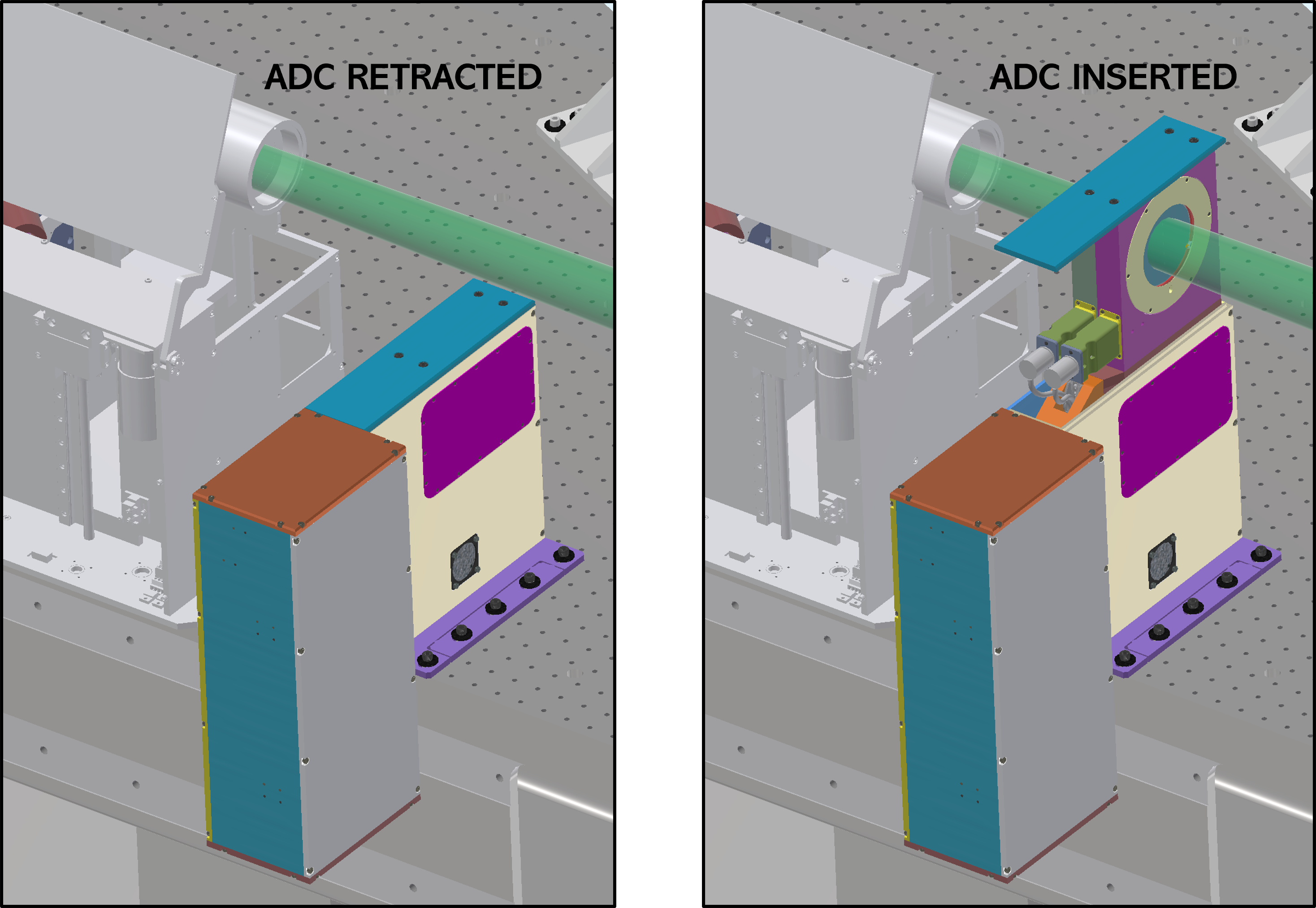}
\vspace{5pt}
\caption{\footnotesize
Deployment states of the RADC integrated into the HJST coud\'e optical path. (\textit{Left}) Retracted configuration, in which the ADC assembly is positioned below the optical axis to completely clear the telescope beam and preserve the full guider field of view. (\textit{Right}) Deployed configuration, where the ADC is inserted into the optical path to provide atmospheric dispersion correction prior to the slit. The insertion mechanism provides rapid deployment while maintaining high positional repeatability and alignment accuracy. 
}
\label{fig:radc_deployment}
\end{figure}

\begin{figure*}[t]
\centering
\includegraphics[width=\textwidth]{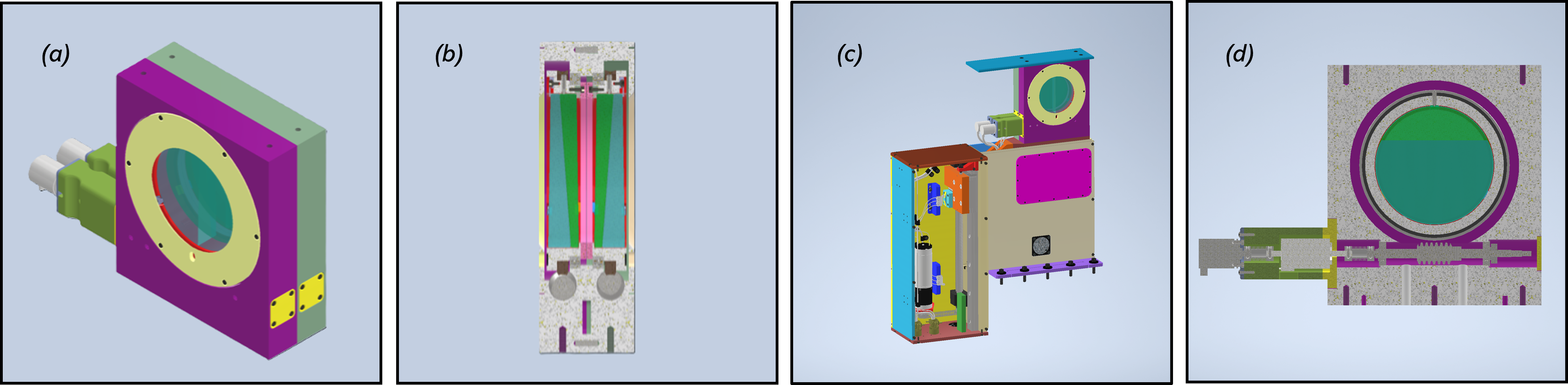}
\vspace{0.1pt}
\caption{\footnotesize Mechanical implementation of the deployable RADC for the Tull Spectrograph. (a) Integrated prism assembly comprising the pair of independently rotating Amici prisms and their supporting structure.(b) Cross-sectional view of the prism rotation stage showing the prism mounting arrangement, bearing support, and internal packaging of the rotation mechanism. (c) Complete RADC assembly incorporating the dual prism rotation stages, protective housing, linear rail-guided insertion mechanism, lead-screw drive, and service enclosure.(d) Detailed cross-sectional view of the worm-drive rotation system used to provide independent prism rotation through a stepper-motor-driven gear train. The combination of independent prism rotation and a deployable insertion stage enables continuous atmospheric-dispersion correction while maintaining compatibility with the existing HJST coud\'e relay infrastructure.
}
\label{fig:radc_mechanical}
\end{figure*}

\noindent The system incorporates three independent motion-control axes consisting of a linear insertion stage and two independently driven prism rotation stages. The insertion mechanism allows the ADC to be deployed during spectroscopic observations and retracted when the full 6\,arcmin guider field of view is required for acquisition. In the retracted configuration, the top surface of the ADC assembly is positioned approximately 100\,mm below the optical axis, fully clearing the 6 arcmin telescope beam. The deployed and retracted operating states are shown in Figure~\ref{fig:radc_deployment}. The insertion stage is designed to provide rapid deployment while maintaining accurate and repeatable positioning of the prism assembly within the optical path.
The insertion stage is supported by a precision linear rail and carriage assembly and is driven by a 12 V, 1.3 A geared servo motor\footnote{\url{https://anaheimautomation.com/bdpg-38-86-12v-3000-r5-2.html}}. Motion is transmitted through a 5/8"-8 stainless-steel lead screw\footnote{\url{https://www.mcmaster.com/98980A140/}} and bronze nut\footnote{\url{https://www.mcmaster.com/1343K132/}}, providing a total insertion stroke of approximately 171.5\,mm. Owing to the relatively long travel required for deployment and retraction, the linear rail, lead screw, and associated drive components are located hanging down the west side of the optical bench, as shown in Figure~\ref{fig:radc_deployment}. This arrangement accommodates the required stroke while maintaining clearance from the optical beam path and existing coud\'e preslit subsystems. The lead screw is supported by a fixed-floating bearing arrangement that accommodates thermal expansion while maintaining positional accuracy. Mechanical hard stops and limit switches define the travel limits, while a calibrated locating surface establishes the deployed position. The mechanism is designed to achieve full deployment in less than 7\,s while maintaining high positional repeatability. The deployed position of the prism assembly is defined by a precision adjustable hard stop  that is set then locked during assembly to position the ADC optical axis on the telescope's optical axis. Limit switches located at both ends of travel define the insertion range and provide over-travel protection.
 Fastened to the lead screw nut, a spring-loaded compliant sub-assembly in the lift ensures that the locating hard point is reached before the insertion limit switch is reached to stop the lift motor. 
 
 \noindent Each Amici prism is housed within an independent rotation stage, shown in Figures~\ref{fig:radc_mechanical}(a) and \ref{fig:radc_mechanical}(b). The prism cells are supported by SKF Reali-Slim\footnote{\url{https://www.skf.com/group/products/thin-section-bearings/reali-slim-thin-section-bearings}} four-point-contact thin-section bearing with 101.6 mm outer diameter and a thickness of 6.35 mm, enabling compact packaging within the constrained relay envelope. The prisms are mounted within precision-machined rotating cells and constrained using dedicated compliant radial and axial locating features to ensure repeatable positioning while minimizing mechanical stress on the optical elements. Independent prism rotation is achieved using worm-gear transmissions driven by 24\,V, 0.6\,A NEMA-08 stepper motors equipped with BiSS-format encoders for closed-loop positional feedback. The complete deployable assembly is shown in Figure~\ref{fig:radc_mechanical}(c). A detailed section through the prism rotation mechanism is presented in Figure~\ref{fig:radc_mechanical}(d). The worm-drive\footnote{\url{https://shop.sdp-si.com/a-1b-6-n24100.html}} architecture provides high angular resolution, positioning repeatability, and resistance to back-driving. Independent control of the two prism stages allows the relative prism angle to be adjusted as a function of telescope zenith distance and coude field rotation, providing the atmospheric-dispersion correction required for broadband observations. The complete moving assembly has a mass of approximately 5.4 kg and incorporates cable-management hardware to accommodate the full insertion stroke. Knife-edge baffles are incorporated near the prism apertures to suppress stray light, while all structural and internal surfaces visible to the beam are specified with broadband blackened finishes. When retracted, the prism assembly is enclosed within a protective housing that minimizes contamination and provides mechanical protection during instrument operations. The protective prism housing and service chamber are connected by a structural strongback that provides the stiffness required to support the insertion mechanism and maintain optical alignment. The complete ADC assembly is fastened to the optical bench and aligned to the telescope optical axis using a three-point adjustment system mounted on the bench. This alignment scheme enables precise positioning of the ADC within the coud\'e optical path while maintaining long-term mechanical stability. The instrument server software queries the Telescope Control System for pointing information to calculate and automatically control the prism rotation angles.

\section{CONCLUSION AND FUTURE WORK}
We have presented the optical and mechanical design of a Rotational Atmospheric Dispersion Corrector (RADC) for the Harlan J. Smith Telescope (HJST) Tull Spectrograph, together with a conceptual K-mirror image rotator for maintaining image orientation during observations of extended and confused-field targets. This upgrade was motivated by the need to improve broadband spectroscopic performance while preserving compatibility with the existing coud\'e optical path and spectrograph infrastructure. The limited mechanical envelope available within the relay imposed significant constraints on both the optical and mechanical implementations. These constraints motivated the selection of a rotational ADC architecture over a longitudinal ADC, as the counter-rotating prism configuration provides continuous atmospheric-dispersion correction without requiring large translational motion or substantial modification of the existing beam path. 
Optical analysis demonstrates that atmospheric dispersion can significantly exceed the projected slit width at large zenith angles, resulting in substantial wavelength-dependent throughput losses. The optimized RADC reduces the residual chromatic displacement to below 0.06\,arcsec over the 350--1000 nm wavelength range and throughout the operational zenith-angle range. 
A deployable opto-mechanical implementation was developed for the RADC to satisfy the space constraints of the HJST coud\'e relay while preserving existing observing capabilities. The design incorporates independent prism-rotation stages and an insertion mechanism that enables atmospheric-dispersion correction to be applied only when required. 
The HJST coude feed exhibits field rotation, and so maintaining a fixed slit position angle remains desirable for observations of extended and non-point-like targets. An optical study confirmed that a reflective K-mirror configuration can provide broadband image rotation without introducing chromatic aberrations or additional refractive material into the optical path. Together, these developments provide a pathway toward improved throughput, reduced chromatic slit losses, enhanced observing efficiency, and improved radial velocity stability for high-resolution spectroscopy with the HJST.

\noindent \textbf{Future Work} : Future efforts will focus on fabrication, assembly, and laboratory validation of the RADC subsystem. Detailed alignment procedures and calibration strategies will be developed to verify prism positioning accuracy, repeatability, and long-term mechanical stability. Laboratory characterization will include measurements of image quality, throughput, residual dispersion, and insertion repeatability prior to installation on the telescope. Following laboratory validation, the RADC will be integrated into the HJST coud\'e optical path and commissioned on sky. Observational testing will evaluate throughput improvements, slit illumination stability, and residual atmospheric-dispersion correction performance over the full operational zenith-angle range. Integration of the control software with the Telescope Control System (TCS) will enable automatic calculation and application of prism rotation angles based on telescope pointing geometry. In parallel, further investigation of the K-mirror image rotator will include detailed opto-mechanical design, tolerance analysis, polarization assessment, and operational studies to evaluate its suitability as a future upgrade for extended-object spectroscopy. 

\appendix    


\bibliography{report} 
\bibliographystyle{spiebib} 

\end{document}